\def\@cite#1#2{$^{\mbox{\scriptsize #1\if@tempswa , #2\fi}}$}

\def\@maketitle{%
  \newpage\spacing{1}\setlength{\parskip}{12pt}%
    {\Large\bfseries\noindent\sloppy \textsf{\@title} \par}%
    {\noindent\sloppy \@author}%
}

\newenvironment{affiliations}{%
    \setcounter{enumi}{1}%
    \setlength{\parindent}{0in}%
    \slshape\sloppy%
    \begin{list}{\upshape$^{\arabic{enumi}}$}{%
        \usecounter{enumi}%
        \setlength{\leftmargin}{0in}%
        \setlength{\topsep}{0in}%
        \setlength{\labelsep}{0in}%
        \setlength{\labelwidth}{0in}%
        \setlength{\listparindent}{0in}%
        \setlength{\itemsep}{0ex}%
        \setlength{\parsep}{0in}%
        }
    }{\end{list}\par\vspace{12pt}}

\newenvironment{addendum}{%
    \setlength{\parindent}{0in}%
    \small%
    \begin{list}{Acknowledgements}{%
        \setlength{\leftmargin}{0in}%
        \setlength{\listparindent}{0in}%
        \setlength{\labelsep}{0em}%
        \setlength{\labelwidth}{0in}%
        \setlength{\itemsep}{12pt}%
        }
    }
    {\end{list}\normalsize}

\newcommand{\hide}[1]{} 

\documentclass[12pt]{article}
\usepackage[super]{cite}
\usepackage{times}
\usepackage{fullpage}

\usepackage{amssymb,amsmath}
\usepackage{graphicx}
\usepackage{subfig}

\begin{document}

\renewcommand{\baselinestretch}{2}\large\normalsize


\renewenvironment{abstract}{
    \setlength{\parindent}{0in}
    \setlength{\parskip}{0in}
    \bfseries
    }{\par\vspace{-6pt}}

\renewcommand\refname{\vspace{-48pt}\setlength{\parskip}{12pt}}



\title{A low mass for Mars from Jupiter’s early gas-driven migration}

\author{Kevin J. Walsh$^{1,2}$, Alessandro Morbidelli$^1$,  Sean N. Raymond$^{3,4}$,\\ David P. O'Brien$^{5}$ \& Avi M. Mandell$^{6}$}

\maketitle

\begin{affiliations}
 \item Universit{\'e} de Nice -- Sophia Antipolis, CNRS, Observatoire de la C\^ote d'Azur, B.P. 4229, 06304 Nice Cedex 4, France
\item Department of Space Studies, Southwest Research Institute, 1050 Walnut Street, Suite 400, Boulder, Colorado, 80302, USA 
\item Universit{\'e} de Bordeaux, Observatoire Aquitain des Sciences de l'Univers, 2 Rue de l'Observatoire, BP 89, F-33270 Floirac Cedex, France
\item CNRS, UMR 5804, Laboratoire d'Astrophysique de Bordeaux, 2 Rue de l'Observatoire, BP 89, F-33270 Floirac Cedex, France
\item Planetary Science Institute, 1700 East Fort Lowell, Suite 106, Tucson, AZ 85719, USA
\item NASA Goddard Space Flight Center, Code 693, Greenbelt, MD 20771, USA
\end{affiliations}

\begin{abstract}

  Jupiter and Saturn formed in a few Myr\cite{Haisch2001} from a
  gas-dominated protoplanetary disk and were susceptible to
  disk-driven migration on timescales of only
  $\sim$100~Kyr\cite{Armitage2007}.  Hydrodynamical simulations show
  that these giant planets can undergo a two-stage,
  inward-then-outward,
  migration\cite{Masset2001,Morby2007,Pierens2008}.  The terrestrial
  planets finished accreting much later\cite{Kleine09}, and their
  characteristics, including Mars' small mass, are best reproduced
  starting from a planetesimal disk with an outer edge at $\sim$1
  AU\cite{Wether1978,Hansen2009}.  Here we present simulations of the
  early Solar System that show how the inward migration of Jupiter to
  1.5 AU, and its subsequent outward migration, leads to a
  planetesimal disk truncated at 1 AU, from which the terrestrial
  planets form over the next 30--50 million years, with a correct
  Earth/Mars mass ratio.  Scattering by Jupiter initially empties, but
  then repopulates the asteroid belt, with inner-belt bodies
  originating between 1--3 AU and outer belt bodies originating
  between and beyond the giant planets.  This explains the significant
  compositional differences across the asteroid belt.  The key aspect
  missing from previous models of terrestrial planet formation is an
  inward, and subsequent outward, migration of Jupiter.  We conclude
  that the behaviour of our giant planets, characterized by
  substantial radial migration, is more similar to that inferred for
  extra-solar planets than previously thought.

\end{abstract}

Hydrodynamic simulations show that isolated giant planets embedded in
gaseous protoplanetary disks carve annular gaps and migrate
inward\cite{Lin1986}.  Saturn migrates faster than Jupiter; if Saturn
is caught in the 2:3 mean motion resonance with Jupiter (see SI,
Section 3, on conditions for this to happen), where their orbital
period ratio is 3/2, generically the two planets start to migrate
outwards until the disappearance of the
disk\cite{Masset2001,Morby2007,Pierens2008,Crida2009ApJ}.  Jupiter
could have migrated inward only before Saturn approached its final
mass and was captured in resonance.  The extents of the inward and
outward migrations are unknown a priori due to uncertainties in the
disk properties and relative timescales for Jupiter and Saturn's
growths.  Thus we search for constraints on where Jupiter's migration
may have reversed or ``tacked''.

The terrestrial planets are best reproduced when the disk of
planetesimals from which they form is truncated with an outer edge at
1~AU (refs 7,8\nocite{Wether1978,Hansen2009}). These conditions are created
naturally if Jupiter tacked at $\sim$~1.5~AU.  However, before
concluding that Jupiter tacked at this distance a major question needs
to be addressed: can the asteroid belt, between 2--3.2 AU, survive the
passage of Jupiter?

Volatile-poor asteroids (mostly S-types) predominate the inner
asteroid belt, while volatile-rich asteroids (mostly C-types)
predominate the outer belt.  These two main classes of asteroids have
partially-overlapping semimajor axis
distributions\cite{Gradie1982,Mothe2003}, though C-types outnumber
S-types beyond $\sim$2.8~AU. We ran a suite of dynamical simulations
to investigate whether this giant planet migration scenario is
consistent with the existence and structure of the asteroid belt. Due
to the many unknowns about giant planet growth and early dynamical
evolution we present a simple scenario reflecting one plausible
history for the giant planets (illustrated in Fig~1).  We provide an
exploration of parameter space in the Supplementary Information that
embraces a large range of possibilities and demonstrates the
robustness of the results.  In all simulations we maintain the
fundamental assumption that Jupiter tacked at 1.5~AU. 

Figure 2 shows how the planet migration affects the small bodies.  The
disk interior to Jupiter is 3.7 Earth-mass ($M_\oplus$) equally
distributed between embryos and planetesimals, while the planetesimal
population exterior to Jupiter is partitioned among inter-planetary
belts and a trans-Neptunian disk (8--13~AU).  The planetesimals from
the inner disk are considered ``S-type'' and those from the outer
regions ``C-type''.  The computation of gas drag assumes 100~km
planetesimals and uses a radial gas density profile taken directly
from hydrodynamic simulations\cite{Morby2007} (see the Supplementary
Information for a more detailed description).

The inward migration of the giant planets shepherds much of the S-type
material inward by resonant trapping, eccentricity excitation and gas
drag. The mass of the disk inside 1~AU doubles, reaching
$\sim$2$M_\oplus$.  This reshaped inner disk constitutes the initial
conditions for terrestrial planet formation.  However, a fraction of
the inner disk ($\sim$14\%) is scattered outward, ending beyond
3~AU. During the subsequent outward migration of the giant planets,
this scattered disk of S-type material is encountered again. Of this
material, a small fraction ($\sim$0.5\%) is scattered inward and left
decoupled from Jupiter in the asteroid belt region as the planets
migrate away.  The giant planets then encounter the material in the
Jupiter-Neptune formation region, some of which ($\sim$0.5\%) is also
scattered into the asteroid belt. Finally, the giant planets encounter
the disk of material beyond Neptune (within 13 AU) of which only
$\sim$~0.025\% reaches a final orbit in the asteroid belt.  When the
giant planets have finished their migration, the asteroid belt
population is in place, while the terrestrial planets require
$\sim$~30 Myr to complete their accretion.

The implanted asteroid belt is composed of two separate populations:
the S-type bodies originally from within 3.0~AU, and the C-types from
between the giant planets and from 8.0-13.0~AU. The actual asteroid
belt consists of more than just S- and C-type asteroids, but this
diversity is expected to result from compositional gradients within
each parent population (see Supplementary Information).  There is a
correlation between the initial and final locations of implanted
asteroids (Fig.~3a). Thus, S-type objects dominate in the inner belt
while C-type objects dominate in the outer belt (Fig.~3b).  Both types
of asteroids share similar distributions in eccentricity and
inclination (Fig.~3c,d).  The asteroid belt is expected to have
eccentricities and inclinations reshuffled during the so-called Late
Heavy Bombardment (LHB)\cite{Gomes2005,Morby2010AJaccept}; the final
orbital distribution in our simulations match the conditions required
by LHB models.

Given the overall efficiency of implantation of $\sim$0.07\%, our
model yields $\sim$1.3$\times 10^{-3}$~$M_\oplus$ of S-type asteroids
at the time of the dissipation of the solar nebula. In the subsequent
4.5 Gyr, this population will be depleted by 50--90\% during the LHB
event\cite{Gomes2005,Morby2010AJaccept} and a further factor of
$\sim$2--3 by chaotic diffusion\cite{Minton2010}. The current-day
asteroid belt is estimated to have 6$\times 10^{-4}$~$M_\oplus$, of
which 1/4 is S-type and 3/4 is C-type\cite{Mothe2003}. Thus our result
is consistent within a factor of a few with the S-type portion of the
asteroid belt. 

The C-type share of the asteroid belt is determined by the total mass
of planetesimals between the giant planets, and between 8--13~AU,
which are not known a priori. Requiring that the mass of implanted
C-type material is 3 times that of the S-type, and given the
implantation efficiencies reported above, this implies
$\sim$0.8~$M_\oplus$ of material between the giant planets is left
over from the giant planet accretion process, or $\sim$16~$M_\oplus$
of planetesimals from the 8.0--13 AU region, or some combination of
the two.

The simulations also found C-type material placed onto orbits crossing
the still-forming terrestrial planets.  For every C-type planetesimal
from beyond 8 AU that was implanted in the outer asteroid belt, 11--28
C-type planetesimals ended up on high-eccentricity orbits that enter
the terrestrial planet forming region (q $<$ 1.0--1.5~AU; see Fig 3),
and may represent a source of water for Earth\cite{Morby2000}.  For
the Jupiter-Uranus region this ratio is 15--20 and for the
Uranus-Neptune region it is 8--15. Thus, depending on which region
dominated the implantation of C-type asteroids, we expect that
3--11$\times$10$^{-2}$~$M_\oplus$ of C-type material entered the
terrestrial planet region. This exceeds by a factor of 6--22 the
minimal mass required to bring the current amount of water to the
Earth ($\sim$5$\times$10$^{-4}$ M$_\oplus$\cite{Lecuyer1997}),
assuming that C-type planetesimals are 10\% water by
mass\cite{Abe2000} .

Concerning the terrestrial planets, the migration of Jupiter creates a
truncated inner disk matching initial conditions of previously
successful simulations\cite{Hansen2009}, though there is a slight
buildup of dynamically excited embryos at 1.0~AU.  Thus, we ran
simulations of the accretion of the surviving objects for 150~Myr.
Earth and Venus grow within the 0.7--1 AU annulus, accreting most of
the mass, while Mars is formed from embryos scattered out beyond the
edge of the truncated disk.  Our final planet mass vs. distance
distribution quantitatively reproduces the large mass ratio between
Earth and Mars and also matches quantitative metrics of orbital
excitation (Fig. 4).

Similar qualitatitive and quantitative results were found for a number
of migration schemes, a range of migration and gas disk dissipation
timescales, levels of gas density and planetesimal sizes (all
described in the Supplementary Information).  This scenario represents
a paradigm shift in the understanding of the early evolution of the
inner solar system. Here C-type asteroids form between and beyond the
giant planets, closer to comets than to S-type asteroids. This can
explain the vast physical differences between S-type and C-type
asteroids, and the physical similarities between the latter and the
comets (as shown by Stardust and micrometeorite
samples\cite{Brownlee2006,Gounelle2008}; see Supplementary Information
for more on physical properties).


If Jupiter and Saturn migrated substantially, then their birth region
could be closer to the estimated location of the snow line,
$\sim$3~AU\cite{Ciesla2006}, rather than beyond 5~AU.  Substantial
migration also points to similarities with observed extra-solar
planetary systems for which migration is seemingly ubiquitous with
giants commonly found at $\sim$1.5~AU\cite{Armitage2007,Butler2006}.
However, the difference between our solar system and the currently
known extra-solar systems is that Jupiter "tacked" at 1.5~AU to
migrate outward due to the presence of Saturn.

\begin{addendum}
\item K.J.W. and A.M. thank the Helmholtz Alliances "Planetary
  Evolution and Life" for financial support. S.N.R and A.M. thank
  CNRS's EPOV and PNP program for funding. D.P.O thanks the NASA PG\&G
  program. A.M.M. thanks the NASA Post-doctoral Program and the
  Goddard Center for Astrobiology. Thanks to the Isaac Newton
  Institute DDP program for hosting some of us at the initial stage of
  the project. Thanks to John Chambers for comments which improved the
  text, and two anonymous referees. Computations were done on the
  CRIMSON Beowulf cluster at OCA.
\item[Author Contribution] K.J.W. managed the simulations and analysis
  and was the primary writer of the manuscript. A.M. initiated the
  project, updated and tested software, ran and analyzed simulations,
  and wrote significant parts of the manuscript. S.N.R helped initiate
  the project, advised on simulations and contributed heavily to the
  manuscript. D.P.O. helped initiate the project and assisted in
  writing. A.M.M. assisted in software updates and in writing.
 \item[Correspondence] Correspondence and requests for materials
   should be addressed to K.J.W.~(email: kwalsh@boulder.swri.edu).
\end{addendum}

\clearpage
\thispagestyle{empty}

\begin{figure}[p]
\centering
\includegraphics*[width=89mm,angle=0]{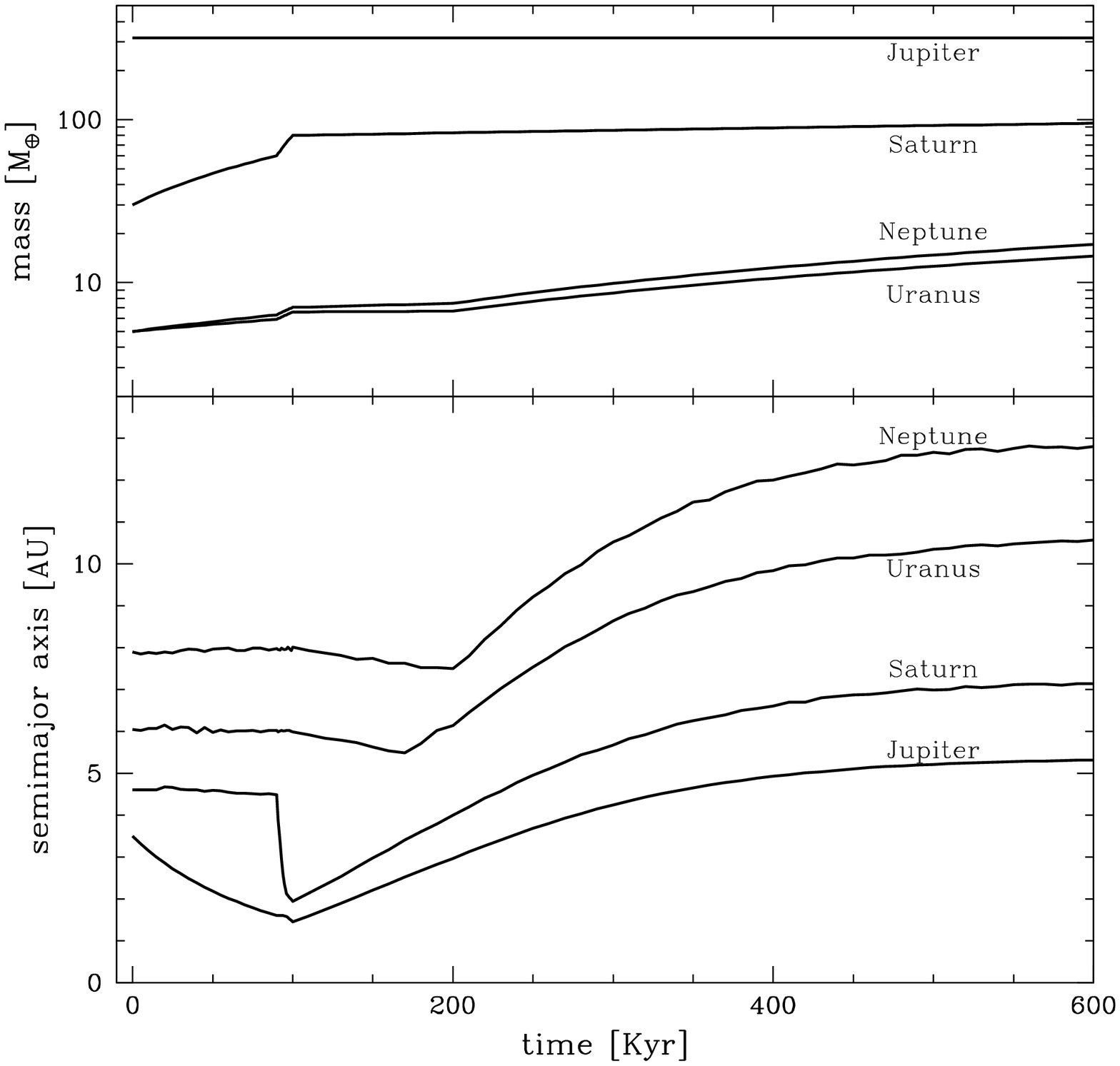}
  \caption{The radial migration and mass growth imposed on the giant
    planets in the reference simulation.  A fully-formed Jupiter
    starts at 3.5 AU, a location expected to be highly favorable for
    giant planet formation due to the presence of the so-called
    snow-line\cite{Ciesla2006}.  Saturn's 30~$M_\oplus$ core is
    initially at $\sim$4.5 AU and grows to 60~$M_\oplus$ as Jupiter
    migrates inward, over $10^5$ years.  Inward type-I migration of
    planetary cores is inhibited in disks with a realistic cooling
    timescale \cite{Paard2006,Paard2008,Kley2008,Masset2009}; thus,
    Saturn's core remains at 4.5 AU during this phase.  Similarly, the
    cores of Uranus and Neptune begin at $\sim$ 6 and 8 AU and grow
    from 5 M$_{\oplus}$, without migrating.  Once Saturn reaches
    60~$M_\oplus$ its inward migration begins\cite{Kley2008}, and is
    much faster than that of the fully grown
    Jupiter\cite{Masset2003}. Thus, upon catching Jupiter, Saturn is
    trapped in the 2:3 resonance\cite{Masset2001}.  Here this happens
    when Jupiter is at 1.5 AU.  The direction of migration is then
    reversed and the giant planets migrate outward together. In
    passing, they capture Uranus and Neptune in resonance which are
    then pushed outwards as well.  Saturn, Uranus and Neptune reach
    their full mass at the end of the migration when Jupiter reaches
    5.4 AU. The migration rate decreases exponentially as the gas disk
    dissipates.  The final orbital configuration of the giant planets
    is consistent with their current orbital configuration when their
    later dynamical evolution is
    considered\cite{Morby2007AJ,Batygin2010} (see SI section 3 for
    extended discussion).}
\label{cartoon}
\end{figure}
\clearpage
\thispagestyle{empty}

\hide{In this scenario, the
    extent of Jupiter's inward migration depends on its migration
    speed (related to disk viscosity) and the time needed for Saturn
    to grow and reach the resonance. The extent of outward migration
    is controlled by the dissipation of the gas disk, and the scale
    height of the disk which governs the migration speed (see SI for
    more discussion).  }

\begin{figure}
\centering
\includegraphics*[width=89mm,angle=0]{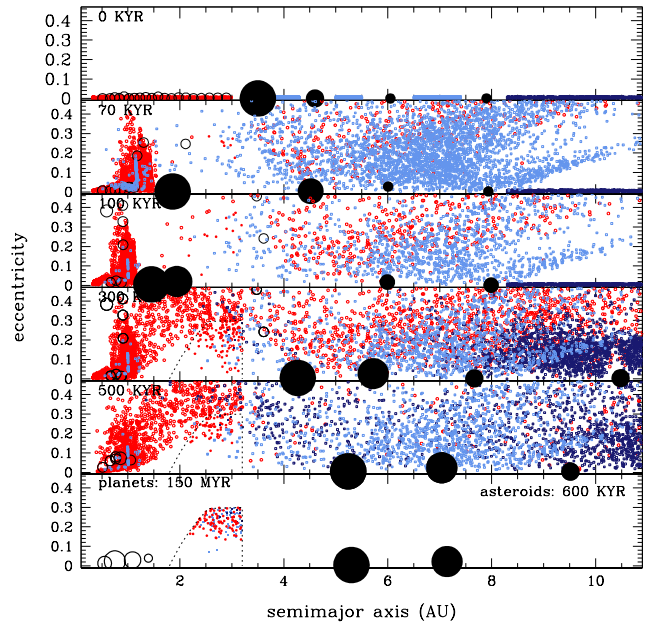}
  \caption{The evolution of the small body populations during giant
    planets growth and migration as described in Fig. 1.  Jupiter,
    Saturn, Uranus and Neptune are represented by large black filled
    circles with evident inward-then-outward migration, and growth of
    Saturn, Uranus and Neptune. S-type planetesimals are represented
    by red dots, initially located between 0.3--3.0 AU. Planetary
    embryos are represented by large open circles scaled by M$^{1/3}$
    (but not in scale relative to the giant planets). The C-type
    planetesimals starting between the giant planets are light blue
    dots and the outer disk planetesimals are dark blue dots,
    initially between 8.0--13.0 AU.  For all planetesimals, filled
    dots are used if they are inside the main asteroid belt and
    smaller open dots otherwise.  The approximate boundaries of the
    main belt are drawn with dotted curves.  The bottom panel combines
    the end state of the giant planet migration simulation (including
    only those planetesimals that finish asteroid belt) with the
    results of simulations of inner disk material ($a <$ 2) evolved
    for 150~Myr (see Fig. 4), reproducing successful terrestrial
    planet simulations\cite{Hansen2009}.}
\label{snapshot}
\end{figure}
\clearpage

\thispagestyle{empty}

\begin{figure}
\centering
\captionsetup{labelformat=empty}
\includegraphics*[width=89mm,angle=0]{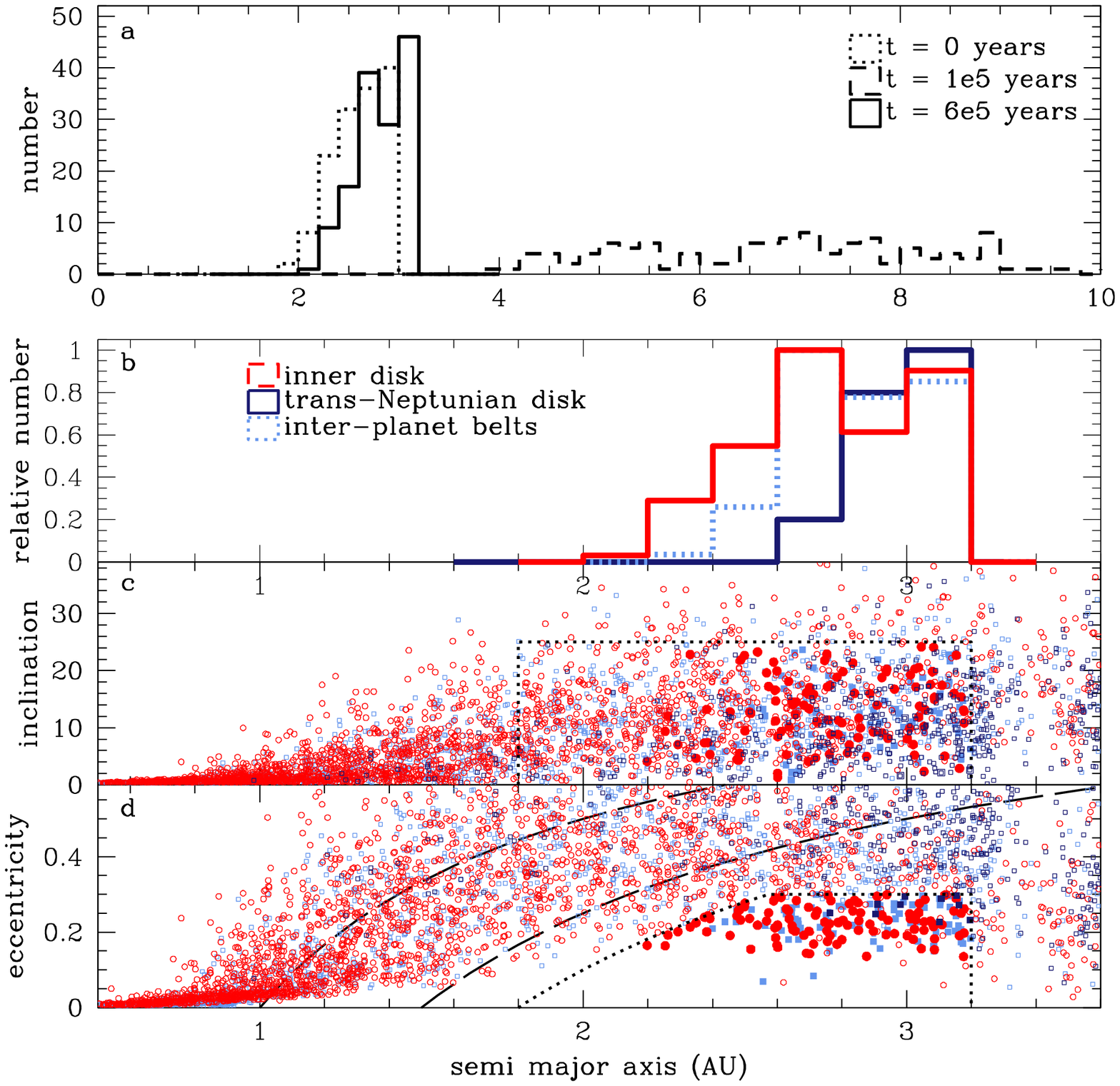}
\caption{Distributions of 100~km planetesimals at the end of giant planet
  migration. Panel a: The semimajor axis distribution for the bodies of the
  inner disk that are implanted in the asteroid belt are plotted at
  three times: the beginning of the simulation (dotted histogram), at
  the end of inward planet migration (dashed) and at the end of
  outward migration (solid).  There is a tendency for S-type
  planetesimals to be implanted near their original location.  Thus,
  the outer edge of their final distribution is related to the
  original outer edge of the S-type disk, which in turn is related to
  the initial location of Jupiter.  The final relative numbers of the
  S-type (red histogram, panel b), the inter-planet population (light
  blue) and the outer-disk (dark blue) planetesimals that are
  implanted in the asteroid belt are shown as a function of semi major
  axis. The orbital inclination (panel c) and eccentricity (panel d) are plotted
  as a function of semi major axis, with the same symbols used in
  Fig. 2. The dotted lines show the extent of the asteroid belt region
  for both inclination and eccentricity, and the dashed lines show the
  limits for perihelion less than 1.0 (left line) and 1.5 (right
  line). Most of the outer-disk material on planet-crossing orbits
  have high eccentricity, while many of the objects from between the
  giant planets were scattered earlier and therefore damped to lower
  eccentricity planet-crossing orbits.}
\label{fig3}
\end{figure}
\clearpage

\thispagestyle{empty}

\begin{figure}
\includegraphics*[width=89mm,angle=0]{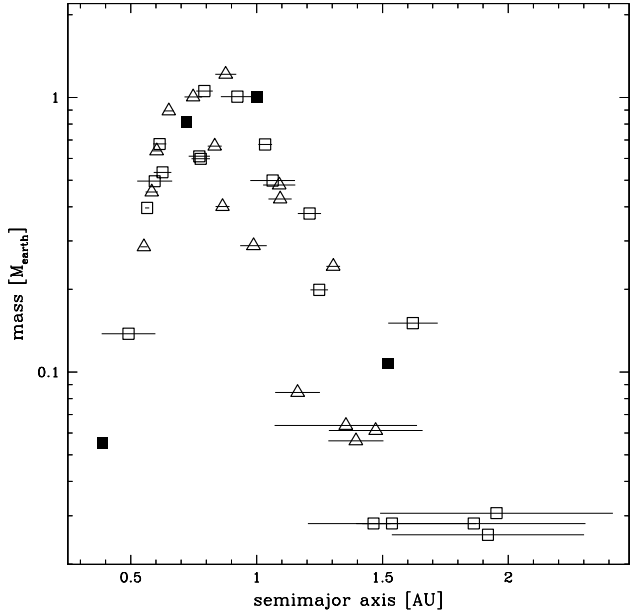}
  \caption{Summary of the 8 terrestrial planet simulations showing the
    mass vs. semi major axis of the synthetic planets (open symbols),
    compared to the real planets (filled squares). The triangles refer
    to simulations starting with 40 embryos of $\sim$ 0.051
    $M_\oplus$, and squares to simulations from 80 embryos of $\sim$
    0.026 $M_\oplus$. The horizontal error bars show the
    perihelion-aphelion excursion of each planet along their
    orbits. The initial planetesimal disk had an inner edge at 0.7 AU
    to replicate previous work\cite{Hansen2009}, and an outer edge at
    $\sim$1.0 AU due to the truncation caused by the inward and
    outward migration of the giant planets as described in the
    text. Half of the original mass of the disk interior of Jupiter
    ($1.85 M_\oplus$) was in $\sim$727 planetesimals. At the end of
    giant planet migration the evolution of all objects inward of 2
    AU was continued for 150 Myr, still accounting for the influence
    from Jupiter and Saturn.  Collisions of embryos with each other
    and with planetesimals were assumed fully accretional.  For this
    set of 8 simulations the average Normalized Angular Momentum
    Deficit\cite{Raymond2009} was 0.0011$\pm$0.0006, as compared to
    0.0018 for the current solar system. Similarly, the Radial Mass
    Concentration\cite{Raymond2009} was 83.8$\pm$12.8 as compared to
    89.9 for the current solar system.}
\label{fig4}
\end{figure}

\thispagestyle{empty}

\end{document}